\pdfoutput=1
\documentclass[12pt]{article}
\usepackage[utf8]{inputenc}
\usepackage{amsmath}
\usepackage{mathtools}
\usepackage{booktabs}
\usepackage{cite}
\usepackage{amssymb}
\usepackage[titletoc,title]{appendix}
\usepackage{mathrsfs}
\usepackage{float}
\usepackage{multirow}
\usepackage{graphicx,caption,subcaption}
\usepackage[space]{grffile}
\usepackage{fullpage}
\usepackage{color}
\usepackage{soul}
\usepackage[dvipsnames]{xcolor}

\definecolor{darkblue}{rgb}{0.1,0.1,.7}
\usepackage[breaklinks=true,linktocpage=true,
colorlinks=true,urlcolor=darkblue,linkcolor=darkblue,
citecolor=darkblue,pdfpagelabels=true,hypertexnames=true,
plainpages=false,naturalnames=false,]{hyperref}

\usepackage{datetime}
\newdateformat{monthyeardate}{%
  \monthname[\THEMONTH], \THEYEAR}
\date{\monthyeardate\today}

\newcommand{\overbar}[1]{\mkern1.5mu\overline{\mkern-1.5mu#1\mkern-1.5mu}\mkern 1.5mu}

\begin{document}

\renewcommand{\arraystretch}{1.3}
\thispagestyle{empty}

{\hbox to\hsize{\vbox{\noindent June 2021}}}

\noindent
\vskip2.0cm
\begin{center}

{\Large\bf Nilpotent superfields for broken abelian symmetries}

\vglue.3in

Yermek Aldabergenov,${}^{a,b,}$\footnote{yermek.a@chula.ac.th} Auttakit Chatrabhuti,${}^{a,}$\footnote{auttakit.c@chula.ac.th} and Hiroshi Isono${}^{a,}$\footnote{hiroshi.isono81@gmail.com}
\vglue.1in

${}^a$~{\it Department of Physics, Faculty of Science, Chulalongkorn University\\
Phayathai Road, Pathumwan, Bangkok 10330, Thailand}\\
${}^b$~{\it Institute of Experimental and Theoretical Physics, Al-Farabi Kazakh National University,
71 Al-Farabi Avenue, Almaty 050040, Kazakhstan}
\vglue.1in

\end{center}

\vglue.3in

\begin{center}
{\Large\bf Abstract}
\vglue.2in
\end{center}

We find new solutions to real cubic constraints on $N=1$ chiral superfields transforming under global abelian symmetries. These solutions describe the low-energy dynamics of a goldstino interacting with an axion (both belonging to the same chiral superfield) with non-linearly realized supersymmetry. We show the relation between our model and the approach of Komargodski and Seiberg for describing goldstino-axion dynamics which uses orthogonal nilpotent superfields.

\newpage

\tableofcontents

\setcounter{footnote}{0}

\section{Introduction}

Spontaneous breaking of global $N=1$ supersymmetry (SUSY) gives rise to a massless Goldstone fermion, or goldstino, which at low energies can be described via the non-linear realization of SUSY first studied by Volkov and Akulov \cite{Volkov:1973ix}. Subsequently, it was realized that non-linear $N=1$ SUSY can also be described in superspace formalism \cite{Rocek:1978nb,Lindstrom:1979kq,Casalbuoni:1988xh,Brignole:1997pe,Komargodski:2009rz,DallAgata:2016syy,Cribiori:2017ngp} (a review of the subject of constrained superfields can be found in Ref. \cite{Farakos:2017bxs}). Other approaches for realising $N=1$ SUSY non-linearly can be found in \cite{Bando:1984mkt,Nagy:2019ywi,Bansal:2020krz}. In particular Komargodski and Seiberg \cite{Komargodski:2009rz} showed that the low-energy non-linear action of the goldstino can be constructed with the help of nilpotent chiral superfields, and in Ref. \cite{Kuzenko:2010ef} Kuzenko and Tyler found non-linear field redefinitions relating Volkov--Akulov and Komargodski--Seiberg models.

In supergravity, constrained superfields have attracted a lot of interest due to their utility in constructing de Sitter vacua even in the absence of scalar fields \cite{Bergshoeff:2015tra,Hasegawa:2015bza,Kuzenko:2015yxa,Kallosh:2015tea,Ferrara:2015gta,Schillo:2015ssx,Bandos:2016xyu,Farakos:2016hly,Cribiori:2016qif,Dudas:2015eha,Antoniadis:2015ala,DallAgata:2015pdd}. The fact that they spontaneously break SUSY and automatically decouple heavy degrees of freedom makes them useful for building inflationary models as well \cite{Antoniadis:2014oya,Ferrara:2014kva,Kallosh:2014via,Kallosh:2014hxa,DallAgata:2014qsj,Kahn:2015mla,Ferrara:2015tyn,Carrasco:2015iij,DallAgata:2015zxp,Dudas:2016eej,Delacretaz:2016nhw,Argurio:2017joe,Aldabergenov:2020atd}.

The minimal goldstino model derived via the constrained superfield approach includes a single chiral superfield $\bf S$ satisfying the quadratic nilpotency constraint ${\bf S}^2=0$.\,\footnote{We adopt the convention of using the bold font for superfields, and the same letters in the regular font for their leading components.} This superfield constraint eliminates the scalar component of $\bf S$, which we call $S$, in terms of the goldstino bilinear $\chi^2$ and introduces higher-derivative terms in the Lagrangian. In the cases where the goldstino is accompanied by other light fields, these fields can be embedded in orthogonal constrained superfields \cite{Komargodski:2009rz,Ferrara:2015tyn,Kahn:2015mla,DallAgata:2016syy,Cribiori:2017ngp}. For example if we have a light Goldstone or a pseudo-Goldstone boson, which we will refer to as an axion, generated by a broken $U(1)$ symmetry, we can embed it in a chiral superfield $\bf T$ satisfying the orthogonality constraint ${\bf S}({\bf T}+\overbar{\bf T})=0$ (in which case the $U(1)$ symmetry transforms $\bf T$ by imaginary shifts), such that the real part of $\bf T$ is eliminated, while its imaginary part can be associated with the axion. From the orthogonality constraint it follows that $({\bf T}+\overbar{\bf T})^3=0$. It is then natural to ask if imposing such a cubic constraint on an independent chiral superfield will yield non-trivial solutions. In this work we attempt to answer this question.

The outline of our paper is as follows. We first recall nilpotent chiral superfields by looking at a simple example in Section \ref{sec_nil_chir}. In Section \ref{sec_cubic} we consider a chiral superfield transforming under a global $U(1)$ symmetry which we assume is spontaneously broken, and impose cubic nilpotency constraints on a $U(1)$-invariant function of the chiral superfield. We solve the cubic constraints in the two cases: where the $U(1)$ acts on the chiral superfield as a shift, and as a phase rotation. The general solutions to the constraints describe non-linear dynamics of a goldstino coupled to an axion, and in Section \ref{sec_eg} we study a minimal Lagrangian as an example. In Section \ref{sec_KS} we review Komargodski--Seiberg (KS) approach for describing goldstino-axion models, and find non-linear field redefinitions relating our construction to the KS approach.

\section{Nilpotent chiral superfield}\label{sec_nil_chir}
The minimal globally supersymmetric Lagrangian for a nilpotent chiral superfield is given by (we use Wess--Bagger \cite{Wess:1992cp} conventions)
\begin{equation}\label{L_minimal}
	{\cal L}=\int\!d^4\theta\,K+\left(\int\!d^2\theta\,W+{\rm h.c.}\right)~,
\end{equation}
where $K$ is the canonical K\"ahler potential and $W$ is a linear superpotential. They have the following expressions
\begin{equation}
	K={\bf S}\overbar{\bf S}~,~~~W=\mu {\bf S}~,\label{KW_minimal}
\end{equation}
where $\mu$ is some constant parameter which we assume is real, and ${\bf S}$ is the chiral superfield. The component expansion of ${\bf S}$ (as a function of the chiral coordinate $y^m=x^m+i\theta\sigma^m\bar{\theta}$) is
\begin{equation}\label{defsfS}
    {\bf S}=S+\sqrt{2}\theta\chi+\theta^2 F~,
\end{equation}
where $S$ is a complex scalar, $\chi$ is a spin-$1/2$ fermion and $F$ is an auxiliary scalar. The chiral superfield ${\bf S}$
satisfies the following nilpotency condition\,\footnote{We can also add a constant $a$ into the constraint, $({\bf S}+a)^2=0$.}
\begin{equation}
    {\bf S}^2=0~.\label{S_nil}
\end{equation}
In terms of the components of ${\bf S}$, the solution to constraint \eqref{S_nil} is $S=\chi^2/(2F)$. This relation eliminates $S$ in terms of the fermion $\chi$, which can now be identified as the goldstino because the corresponding auxiliary field $F$ must be non-vanishing. Hence SUSY is realised non-linearly. 

After expanding Lagrangian \eqref{L_minimal} in terms of the component fields, and using $S=\chi^2/(2F)$, we have
\begin{equation}
	{\cal L}=\frac{\chi^2}{2F}\Box\frac{\bar{\chi}^2}{2\overbar{F}}-i\chi\sigma^m\partial_m\bar{\chi}+\mu(F+\overbar{F})+F\overbar{F}~.
\end{equation}
The auxiliary field $F$ can be integrated out via its equation of motion
\begin{equation}
	F=-\mu-\frac{\bar{\chi}^2}{4\mu^3}\Box\chi^2+\frac{3\chi^2\bar{\chi}^2}{16\mu^7}\Box\chi^2\Box\bar{\chi}^2~,
\end{equation}
where the leading term is the usual contribution from the superpotential, while the fermionic terms come from the solution to the nilpotency constraint. Plugging this back into the Lagrangian yields,
\begin{equation}
	{\cal L}=-i\chi\sigma^m\partial_m\bar{\chi}-\mu^2+\frac{\chi^2}{4\mu^2}\Box\bar{\chi}^2-\frac{\chi^2\bar{\chi}^2}{16\mu^6}\Box\chi^2\Box\bar{\chi}^2\,.
\end{equation}
This minimal Lagrangian describes non-linear dynamics of the goldstino which transforms under SUSY as,
\begin{equation}
    \delta_\epsilon\chi_\alpha=-\mu\epsilon_\alpha-i\mu^{-1}\sigma^m_{\alpha\dot\alpha}\bar{\epsilon}^{\dot\alpha}(\chi\partial_m\chi)+{\cal O}(\chi^2,\bar{\chi}^2)~,
\end{equation}
where $\epsilon$ is the constant transformation parameter of global SUSY.

\section{Real cubic constraints}\label{sec_cubic}

A superfield ${\bf S}$ that satisfies the quadratic chiral constraint \eqref{S_nil} also trivially satisfies the real cubic constraints,
\begin{equation}
	({\bf S}+\overbar{\bf S})^3=0~,~~~{\rm and}~~~({\bf S}\overbar{\bf S})^3=0~,
\end{equation}
where the former is invariant w.r.t. its imaginary shifts, while the latter is invariant w.r.t. phase rotations of ${\bf S}$. Our goal is to find general solutions to this class of cubic constraints, including the solutions that do not satisfy the quadratic constraints of the form ${\bf S}^2=0$. As we are going to show, the most general solutions to the cubic constraints eliminate one real scalar, so that the other one can be used to describe a light axion of a broken global abelian symmetry.

A similar class of cubic constraints was studied by Kuzenko \cite{Kuzenko:2017oni}, where it was applied to the deformed real linear superfield ${\bf L}$ defined by $\overbar{D}^2 {\bf L}=-4m$, where $m$ is a non-vanishing deformation parameter.

\subsection{Shift-symmetric case}\label{subsec_shift}

First we consider a chiral superfield $\bf{S}$, whose components are given in \eqref{defsfS}, transforming by imaginary shifts under a (spontaneously broken) global $U(1)$, i.e. ${\bf S}\rightarrow{\bf S}+i\vartheta$ with a constant transformation parameter $\vartheta$. If this $U(1)$ is an $R$-symmetry, $\chi$ and $F$ must transform accordingly.

Introducing the following invariant quantity,
\begin{equation}
	{\bf\Sigma}\equiv {\bf S}+\overbar{\bf S}-2a~,
\end{equation}
where the (real) constant $a$ is included as a possible VEV of $S$, we impose the following cubic constraint,
\begin{equation}\label{cubic_shift_constr}
	{\bf\Sigma}^3=0~.
\end{equation}
In terms of the regular coordinates $x^m$, $\bf\Sigma$ has the expansion,
\begin{multline}
	{\bf\Sigma}=\Sigma+\sqrt{2}\theta\chi+\sqrt{2}\bar{\theta}\bar{\chi}+i\theta\sigma^m\bar{\theta}\partial_m(S-\overbar{S})\\+\frac{i}{\sqrt{2}}\theta^2\bar{\theta}\overbar{\sigma}^m\partial_m\chi+\frac{i}{\sqrt{2}}\bar{\theta}^2\theta\sigma^m\partial_m\bar{\chi}+\theta^2 F+\bar{\theta}^2\overbar{F}+\frac{1}{4}\theta^2\bar{\theta}^2\Box(S+\overbar{S})~,
\end{multline}
where $\Sigma\equiv S+\bar{S}-2a$.

The leading component of the constraint \eqref{cubic_shift_constr} implies the general form of $\Sigma$,
\begin{equation}
	\Sigma=\chi^2\beta+\bar\chi^2\bar\beta+\chi\sigma^m\bar\chi\,\omega_m~,\label{ansatz}
\end{equation}
where $\beta$ and $\omega_m$ are some functions of $\chi$, $\bar{\chi}$, ${\rm Im}(S)$, and their spacetime derivatives. Let us parametrize $S$ as
\begin{equation}
	S=\phi+i\varphi~,
\end{equation}
so that the real part is related to $\Sigma$ as $\phi=a+\Sigma/2$, and the imaginary part $\varphi$ is the would-be axion of the broken $U(1)$ shift symmetry.

The $\theta^2\bar{\theta}^2$-component of constraint \eqref{cubic_shift_constr} reads,
\begin{equation}
	\left(A+\frac{1}{4}\Sigma\Box\Sigma\right)\Sigma=B~,\label{eq_Sigma_theta^4}
\end{equation}
where
\begin{eqnarray}
	A&\equiv & 2|F|^2-2\partial_m\varphi\partial^m\varphi-i\chi\sigma^m\partial_m\bar{\chi}+i\partial_m\chi\sigma^m\bar{\chi}~,\\
	B&\equiv &\chi^2\overbar F+\bar\chi^2 F+2\chi\sigma^m\bar\chi\partial_m\varphi~.
\end{eqnarray}
Note that $B^2={\cal O}(\chi^2\bar{\chi}^2)$ and $B^3=0$. This equation is similar to what was obtained in \cite{Kuzenko:2017oni} for the deformed real linear superfield $\bf L$, because the superfield ${\bf\Sigma}$ has the same component structure as $\bf L$, except that the auxiliary field $F$ of the chiral superfield $\bf S$ is replaced by the deformation parameter $m$ of the linear superfield, and $\partial_m\varphi$ is replaced by the Hodge-dual of the field strength of a two-form field in $\bf L$. The two-form field can be dualized to a real scalar (axion) in four dimensions, so the field content is the same as that of a chiral superfield.

Eq.~\eqref{eq_Sigma_theta^4} can be solved iteratively (because $\chi^3=0$), and the general solution is,
\begin{equation}
	\Sigma=\frac{B}{A}-\frac{B^2}{4A^4}\Box B~.\label{Sigma_sol}
\end{equation}
After expanding the solution in terms of $\chi$ and $\bar{\chi}$ we arrive at,
\begin{equation}
	\Sigma=\chi^2\beta+\bar{\chi}^2\bar\beta+\frac{2}{U}\chi\sigma^m\bar{\chi}\partial_m\varphi~,\label{Sigma_full}
\end{equation}
where $U\equiv 2(|F|^2-\partial_m\varphi\partial^m\varphi)$ and\,\footnote{We use the convention ${{\sigma^{mn}}_{\alpha}}^{\beta}=\frac{1}{4}\left(\sigma^m_{\alpha\dot{\alpha}}\overbar{\sigma}^{n\dot{\alpha}\beta}-\sigma^n_{\alpha\dot{\alpha}}\overbar{\sigma}^{m\dot{\alpha}\beta}\right)$, and $\epsilon^{0123}=1$.}
\begin{align}
\begin{split}
	\beta\equiv\frac{\overbar{F}}{U}&+\frac{i\bar{\chi}}{U^2}\left(\overbar{F}\overbar{\sigma}^m\partial_m\chi-\partial_m\varphi\partial^m\bar{\chi}+2\partial_m\varphi\overbar{\sigma}^{mn}\partial_n\bar{\chi}\right)\\
	&-\frac{\bar{\chi}^2}{2U^3}\Big[ F\partial_m\bar{\chi}\overbar{\sigma}^{mn}\partial_n\bar{\chi}+\overbar{F}\partial_m\chi\sigma^{mn}\partial_n\chi\\
	&\qquad\qquad+\partial_m\varphi\partial_n\chi(2\sigma^{m}\eta^{nk}-\sigma^{n}\eta^{km}-\sigma^{k}\eta^{mn}-i\epsilon^{mnkl}\sigma_l)\partial_k\bar{\chi}\Big]~.\label{beta_def}
\end{split}
\end{align}
Or, in terms of $\phi$ we have,
\begin{equation}\label{phi_sol}
    \phi=a+\frac{\Sigma}{2}=a+\frac{\chi^2}{2}\beta+\frac{\bar\chi^2}{2}\bar\beta+\frac{1}{U}\chi\sigma^m\bar{\chi}\,\partial_m\varphi~.
\end{equation}
Note that we do not have any restrictions on $a$ and it can be consistently set to zero.

Of course, the solution is valid only if $U= 2(|F|^2-\partial_m\varphi\partial^m\varphi)\neq 0$. Here we assume that $U>0$ for physically interesting models, in which case the fluctuations of $\varphi$ are bounded from above by the SUSY breaking scale, which is not at all surprising since the effective theory is supposed to operate on scales well below it. We verified that the solution \eqref{Sigma_full} is compatible with the rest of the components of the superfield equation \eqref{cubic_shift_constr}.

\medskip

\noindent{\bf Nilpotent chiral superfield as a special case}. Solution \eqref{Sigma_full} can be checked for consistency by additionally imposing the quadratic constraint $({\bf S}-a)^2=0$, in which case it reduces to $S-a=\chi^2/(2F)$ so that
\begin{equation}
	 \Sigma\equiv S+\bar{S}-2a=\frac{\chi^2}{2F}+\frac{\bar{\chi}^2}{2\overbar{F}}~.\label{Sigma_spec} 
\end{equation}
This eliminates $\varphi$ as
\begin{equation}
	\varphi=-\,\frac{i}{2}(S-\bar{S})=-\frac{i}{4}\left(\frac{\chi^2}{F}-\frac{\bar{\chi}^2}{\overbar{F}}\right)~.\label{varphi_elim}
\end{equation}
Inserting solution \eqref{varphi_elim} into the general solution for $\Sigma$ in \eqref{Sigma_full} reproduces exactly Eq.~\eqref{Sigma_spec}, as it should.

\subsection{Phase-symmetric case}\label{subsec_phase}

Suppose a theory is symmetric (or approximately symmetric) w.r.t. $U(1)$ phase rotations of a chiral superfield ${\bf Z}\rightarrow {\bf Z}e^{i\vartheta}$. We denote the components of $\bf Z$ as,
\begin{equation}
    {\bf Z}=Z+\sqrt{2}\theta\,\psi+\theta^2F^z~,
\end{equation}
where the fermion $\psi$ will be identified with the goldstino. Depending on whether the $U(1)$ is an $R$-symmetry or not, $\psi$ and $F^z$ may transform differently from $Z$.

We can now define a $U(1)$ invariant combination,
\begin{equation}
    {\bf \Lambda}\equiv {\bf Z}\overbar{\bf Z}-b^2~,
\end{equation}
where $b$ is some real constant, and impose the cubic constraint,
\begin{equation}
    {\bf \Lambda}^3=0~.\label{cubic_phase_constr}
\end{equation}
It is convenient to parametrize the scalar $Z$ as,
\begin{equation}
    Z=|Z|e^{i\zeta}~,
\end{equation}
where $\zeta$ is the angular scalar (axion), and the VEV of the radial mode is $\langle |Z|\rangle=b$. Again, we assume the $U(1)$ is spontaneously broken.

The $\theta^2\bar{\theta}^2$-component of the constraint \eqref{cubic_phase_constr} reads,
\begin{equation}\label{eq_Lambda_theta^4}
    \left(G+\frac{1}{4}\Box\Lambda\right)\Lambda^2+b^2H\Lambda=b^4I~,
\end{equation}
where $\Lambda\equiv\mathbf{\Lambda}|=Z\bar{Z}-b^2$, and
\begin{align}
	G &\equiv 
    3|F^z|^2-5b^2\partial_m\zeta\partial^m\zeta~,\\
	H &\equiv
	2|F^z|^2-2b^2\partial_m\zeta\partial^m\zeta-i\psi\sigma^m\partial_m\bar{\psi}+i\partial_m\psi\sigma^m\bar{\psi}\nonumber\\
	&\qquad\qquad\qquad-\tfrac{7}{2b}\psi^2\overbar{F}^ze^{-i\zeta}-\tfrac{7}{2b}\bar{\psi}^2F^ze^{i\zeta}-8\psi\sigma^m\bar{\psi}\partial_m\zeta~,\\
	I &\equiv
	\tfrac{1}{b}\psi^2\overbar{F}^z e^{-i\zeta}+\tfrac{1}{b}\bar{\psi}^2F^z e^{i\zeta}+2\psi\sigma^m\bar{\psi}\partial_m\zeta-\tfrac{2}{b^2}\psi^2\bar{\psi}^2~.
\end{align}
Note that $I^2={\cal O}(\psi^2\bar{\psi}^2)$ and $I^3=0$.

In the limit $b=0$ the only meaningful solution is $\Lambda^2=Z^2\overbar{Z}^2=0$ which means $Z^2=0$, and eliminates both real scalar degrees of freedom in terms of the goldstino.

For a general solution to Eq.~\eqref{eq_Lambda_theta^4} we assume $b\neq 0$. In this case \eqref{eq_Lambda_theta^4} is solved by,
\begin{equation}
    \Lambda=\frac{b^2I}{H}-\frac{b^2I^2}{H^3}\left(G+\frac{b^2}{4H}\Box I\right)~,
\end{equation}
or after expanding in terms of $\psi$ and $\bar{\psi}$ we get,
\begin{equation}
    \Lambda=b\left(\psi^2\gamma+\bar{\psi}^2\bar{\gamma}+\frac{2b}{Y}\psi\sigma^m\bar{\psi}\partial_m\zeta+\frac{b}{Y^2}\psi^2\bar{\psi}^2\partial_m\zeta\partial^m\zeta\right)~,\label{Lambda_sol}
\end{equation}
where $Y\equiv 2(|F^z|^2-b^2\partial_m\zeta\partial^m\zeta)$, and
\begin{align}
\begin{split}
    \gamma\equiv\frac{\overbar{F}^z}{Y}e^{-i\zeta}&+\frac{i\bar{\psi}}{Y^2}\left(\overbar{F}^z e^{-i\zeta}\overbar{\sigma}^m\partial_m\psi-b\partial_m\zeta\partial^m\bar{\psi}+2b\partial_m\zeta\overbar{\sigma}^{mn}\partial_n\bar{\psi}\right)\\
    &-\frac{\bar{\psi}^2}{2Y^3}\Big[ F^z e^{i\zeta}\partial_m\bar{\psi}\overbar{\sigma}^{mn}\partial_n\bar{\psi}+\overbar{F}^ze^{-i\zeta}\partial_m\psi\sigma^{mn}\partial_n\psi\\
    &\qquad\qquad+b\partial_m\zeta\partial_n\psi(2\sigma^{m}\eta^{nk}-\sigma^{n}\eta^{km}-\sigma^{k}\eta^{mn}-i\epsilon^{mnkl}\sigma_l)\partial_k\bar{\psi}\Big]~.
\end{split}
\end{align}
We have defined $\gamma$ in a way such that it resembles $\beta$ from Eq.~\eqref{beta_def} in its structure. The solution is well-defined as long as $Y= 2(|F^z|^2-b^2\partial_m\zeta\partial^m\zeta)\neq 0$.

Since $\Lambda=|Z|^2-b^2$, the radial mode of $Z$ can be written with the help of $\Lambda^3=0$ as,
\begin{equation}
    |Z|=\sqrt{b^2+\Lambda}=b+\frac{\Lambda}{2b}-\frac{\Lambda^2}{8b^3}~,
\end{equation}
so that using \eqref{Lambda_sol} we obtain,
\begin{equation}
    |Z|=b+\frac{\psi^2}{2}\gamma+\frac{\bar\psi^2}{2}\bar{\gamma}+\frac{b}{Y}\psi\sigma^m\bar{\psi}\partial_m\zeta-\frac{\psi^2\bar{\psi}^2}{4bY^2}\left(|F^z|^2-3b^2\partial_m\zeta\partial^m\zeta\right)~.\label{z_sol}
\end{equation}

There is an alternative way to derive solution \eqref{z_sol}: we could start from solution \eqref{phi_sol} for the shift-symmetric case, and identify ${\bf S}=\log{\bf Z}$ assuming $\langle {\bf Z}\rangle\neq 0$. Using the same expansion for ${\bf S}$ and ${\bf Z}$ as before,
\begin{equation}
    {\bf S}=S+\sqrt{2}\theta\chi+\theta^2F~,~~~{\bf Z}=Z+\sqrt{2}\theta\psi+\theta^2F^z~,
\end{equation}
the components of ${\bf S}=\log{\bf Z}$ read,
\begin{equation}
    S=\log Z~,~~~\chi=\frac{\psi}{Z}~,~~~F=\frac{F^z}{Z}+\frac{\psi^2}{2Z^2}~.\label{s_z_ident}
\end{equation}
Thus, the real ($\phi$) and imaginary ($\varphi$) parts of $S$ can be written as $\phi=\log|Z|$ and $\varphi=\zeta$ (also notice that the VEV of the first relation above reads $a=\log b$, where $b\neq 0$ while $a$ can be zero). Plugging \eqref{s_z_ident} into solution \eqref{phi_sol} and expanding in powers of $\psi$ and $\bar{\psi}$, we obtain exactly the solution \eqref{z_sol} for $|Z|$. In other words, the solutions to the constraints $({\bf Z}\overbar{\bf Z}-b^2)^3=0$ and $[\log({\bf Z}\overbar{\bf Z}/b^2)]^3=0$ coincide.

\medskip

\noindent{\bf Nilpotent chiral superfield as a special case}. Let us once again verify that applying the appropriate quadratic constraint consistently eliminates the axion. The relevant quadratic constraint in this case is $({\bf Z}-b)^2=0$, and its solution $Z-b=\psi^2/(2F^z)$ can be written in terms of the radial and angular parts as,
\begin{eqnarray}
    |Z|&=&b+\frac{\psi^2}{4F^z}+\frac{\bar{\psi}^2}{4\overbar{F}^z}+\frac{\psi^2\bar{\psi}^2}{16b|F^z|^2}~,\label{z_quad_sol}\\
    \zeta &=&-\frac{i}{2}\log\frac{Z}{\overbar Z}=\frac{i}{4b}\left(\frac{\bar{\psi}^2}{\overbar{F}^z}-\frac{\psi^2}{F^z}\right)~.\label{zeta_quad_sol}
\end{eqnarray}
Plugging $\zeta$ from \eqref{zeta_quad_sol} into Eq.~\eqref{z_sol}, we indeed get Eq.~\eqref{z_quad_sol}.

SUSY transformation laws for the independent fields after solving the cubic constraints can be found in Appendix \ref{App_SUSY} for both shift-symmetric and phase-symmetric cases.

\section{Goldstino-axion interactions}\label{sec_eg}

To demonstrate the Lagrangian and interactions between the goldstino and the axion, let us choose the simplest setup with a single chiral superfield and non-vanishing $F$-term. We fix the K\"ahler potential and superpotential as in Section \ref{sec_nil_chir},
\begin{equation}
	K={\bf S}\overbar{\bf S}~,~~~W=\mu {\bf S}~,\label{KW_minimal2}
\end{equation}
which leads to the component action,
\begin{align}
    {\cal L}&=-\partial_m S\partial^m\overbar S-i\chi\sigma^m\partial_m\bar\chi+\mu(F+\overbar F)+|F|^2~\nonumber\\
    &=\phi\Box\phi+\varphi\Box\varphi-i\chi\sigma^m\partial_m\bar\chi+\mu(F+\overbar F)+|F|^2~,\label{L_eg}
\end{align}
where we again parametrized $S=\phi+i\varphi$.

The Lagrangian is invariant w.r.t. both constant shifts and constant phase transformations of $S$. We choose the shift-symmetry in $\varphi$ direction as the relevant $U(1)$ symmetry, so that $\varphi$ is the axion which is exactly massless in this case (we assume that a microscopic theory spontaneously breaks the $U(1)$ and generates a large mass for $\phi$, which can always be arranged by e.g. considering higher order corrections to the minimal K\"ahler potential \cite{Komargodski:2009rz,Cribiori:2017ngp}). Then, we apply the nilpotency constraint given by Eq.~\eqref{cubic_shift_constr} and its solution in terms of $\phi$ given by Eq.~\eqref{phi_sol}. Substituting this solution into the Lagrangian \eqref{L_eg} and integrating out $F$ and $\overbar F$ we get,
\begin{align}
    {\cal L}=\,\,&\varphi\Box\varphi -i\chi\sigma^m\partial_m\bar\chi-\mu^2~\nonumber\\
    &+\frac{\mu\chi^2+\mu\bar\chi^2-2\chi\sigma^m\bar\chi\partial_m\varphi}{8(\mu^2-\partial_n\varphi\partial^n\varphi)^2}\left[\mu\partial_k\chi\partial^k\chi+\mu\partial_k\bar\chi\partial^k\bar\chi-2\partial_k\varphi\partial_l\chi\sigma^k\partial^l\bar\chi\right]+\ldots~,\label{L_bi}
\end{align}
where the ellipsis stands for the terms trilinear and quadrilinear in $\chi$ and $\bar\chi$. The solution to the $F$-field equation of motion is given in Appendix \ref{Appendix_F_term}, and the full Lagrangian after eliminating $F$ can be seen in Eq.~\eqref{App_L_final}.

The parameter $\mu$ becomes an expectation value of $F$ ($\langle F\rangle=-\mu$) and acts as the order parameter of SUSY breaking. Notice that the interaction term in \eqref{L_bi} includes the factor of $(\mu^2-\partial\varphi\partial\varphi)^{-2}$ which can be expanded to any given order in the derivatives $\partial\varphi\partial\varphi$. Similarly, trilinear and quadrilinear terms contain negative powers of $(\mu^2-\partial\varphi\partial\varphi)$, as can be seen from Eq. \eqref{App_L_final}.

\section{Relation to Komargodski--Seiberg model}\label{sec_KS}

Komargodski and Seiberg \cite{Komargodski:2009rz} showed that when a goldstino is accompanied by an axion at low energies, in addition to the goldstino chiral superfield satisfying ${\bf X}^2=0$, one can introduce a chiral superfield $\bf T$ containing the axion and satisfying the orthogonality constraints,
\begin{align}
    {\bf X}({\bf T}+\overbar{\bf T})&=0~,\label{constr_XTT}\\
    {\bf X}\overbar D_{\dot\alpha}\overbar{\bf T}&=0~,\label{constr_XDT}
\end{align}
where the latter constraint can derived from the former by acting on it with $\overbar D_{\dot\alpha}$. The role of these constraints is to eliminate all the components of ${\bf T}$ except its imaginary part which is identified with the axion. Let us denote the components of ${\bf X}$ and ${\bf T}$ as follows,
\begin{equation}
    {\bf X}=X+\sqrt{2}\theta\lambda+\theta^2F^x~,~~~{\bf T}=T+\sqrt{2}\theta\kappa+\theta^2F^t~,
\end{equation}
where $T=t+i\tau$, with $\tau$ being the axion.

The constraint ${\bf X}^2=0$ sets $X=\lambda^2/(2F^x)$, where $\lambda$ becomes the goldstino. Then the $\theta^2$-component of constraint \eqref{constr_XDT} eliminates $\bar\kappa$ by the relation,
\begin{equation}
    \bar\kappa_{\dot\alpha}=-i\frac{\lambda^\alpha}{F^x}\sigma^m_{\alpha\dot\alpha}\partial_m(t-i\tau)~,
\end{equation}
while the $\theta^2\bar\theta$-component of the same constraint yields,
\begin{equation}
    \overbar F^t=\left[-\frac{\lambda}{F^x}\sigma^m\overbar\sigma^n\partial_m\left(\frac{\lambda}{F^x}\right)\partial_n+\frac{\lambda^2}{2(F^x)^2}\Box\right](t-i\tau)~.\label{Ft_sol}
\end{equation}
Finally, constraint \eqref{constr_XTT} eliminates the saxion $t$ as,
\begin{align}
    t=-\,\frac{\lambda}{2F^x}\sigma^m\frac{\bar\lambda}{\overbar F^x}\partial_m\tau &-\frac{i\bar\lambda^2}{8(\overbar F^x)^2}\frac{\lambda}{F^x}\sigma^n\overbar\sigma^m\partial_n\left(\frac{\lambda}{F^x}\right)\partial_m\tau+
    \frac{i\lambda^2}{8( F^x)^2}\partial_n\left(\frac{\bar\lambda}{\overbar F^x}\right)\overbar\sigma^m\sigma^n\frac{\bar\lambda}{\overbar F^x}\partial_m\tau\nonumber\\
    &-\frac{\lambda^2\bar\lambda^2}{16|F^x|^4}\partial_m\left(\frac{\bar\lambda}{\overbar F^x}\right)\left(\overbar\sigma^k\eta^{mn}+i\epsilon^{kmnl}\overbar\sigma_l\right)\partial_n\left(\frac{\lambda}{F^x}\right)\partial_k\tau~.\label{t_sol}
\end{align}

It was shown in \cite{Komargodski:2009rz} that if the two constrained chiral superfields ${\bf X}$ and ${\bf T}$ transform under a global $U(1)$ as,
\begin{equation}
    {\bf X}\rightarrow{\bf X}e^{i\vartheta}~,~~~{\bf T}\rightarrow {\bf T}+i\vartheta~,
\end{equation}
they can form a $U(1)$-invariant combination satisfying the following (secondary) cubic constraint,
\begin{equation}\label{P_def}
    {\bf P}\equiv \alpha{\bf X}+e^{\bf T}~,~~~\left(|{\bf P}|^2-1\right)^3=0~,
\end{equation}
where $\alpha$ is some complex parameter.

Alternatively one can take the logarithm of ${\bf P}$, 
\begin{equation}
    {\bf Q}\equiv\log\left(\alpha{\bf X}+e^{\bf T}\right)~,
\end{equation}
which satisfies the shift-symmetric constraint\,\footnote{Recall that $(|{\bf P}|^2-1)^3=0$ and $[\log\,({\bf P}\overbar{\bf P})]^3=0$ lead to the same solution, so using ${\bf P}=e^{\bf Q}$ in the latter we can obtain shift-symmetric constraint \eqref{constr_Q}.},
\begin{equation}
    \left({\bf Q}+\overbar{\bf Q}\right)^3=0~,\label{constr_Q}
\end{equation}
where ${\bf Q}$ transforms by imaginary shifts under the $U(1)$ transformation. Constraint \eqref{constr_Q} is exactly of the form given by Eq.~\eqref{cubic_shift_constr} which we solved in Subsection \ref{subsec_shift} (here we ignore the constant $a$). Therefore we can identify ${\bf Q}={\bf S}$ and express the independent components of $\bf S$ (which are $\varphi$, $\chi$, and $F$) in terms of the independent components of $\bf X$ and $\bf T$ (which are $\tau$, $\lambda$, and $F^x$),
\begin{align}
    \varphi &=\tau-\frac{i}{4}\left(\frac{\alpha\lambda^2}{F^x}e^{-i\tau}-\frac{\bar\alpha\bar\lambda^2}{\overbar F^x}e^{i\tau}\right)~,\label{varphi_tau}\\
    \chi_\alpha &=\alpha\lambda_\alpha e^{-t-i\tau}+i\sigma^m_{\alpha\dot\alpha}\frac{\bar\lambda^{\dot\alpha}}{\overbar F^x}\partial_m(t+i\tau)\left(1-\frac{\alpha\lambda^2}{2F^x}e^{-i\tau}\right)~,\label{chi_lambda}\\
    F &=\alpha e^{-t-i\tau}F^x\Bigg\{1+i\frac{\lambda}{F^x}\sigma^m\frac{\bar\lambda}{\overbar F^x}\partial_m(t+i\tau)-\frac{\lambda^2\bar\lambda^2}{4|F^x|^4}\partial^m\tau\partial_m\tau\Bigg\}\nonumber\\
    & \qquad -\left(1-\frac{\alpha\lambda^2}{2F^x}e^{-i\tau}\right)\left[\partial_m\left(\frac{\bar\lambda}{\overbar F^x}\right)\overbar\sigma^n\sigma^m\frac{\bar\lambda}{\overbar F^x}\partial_n-\frac{\bar\lambda^2}{2(\overbar F^x)^2}\Box\right](t+i\tau)~,\label{F_Fx}
\end{align}
where $t$ is given by Eq.~\eqref{t_sol}, and we used the fact that $e^{-t}=1+{\cal O}(\lambda\sigma^m\bar\lambda)$. Eqs.~\eqref{varphi_tau}, \eqref{chi_lambda}, and \eqref{F_Fx} make up the nonlinear field redefinitions relating our model of Subsection \ref{subsec_shift} to the Komargodski--Seiberg model \cite{Komargodski:2009rz} of goldstino-axion dynamics. Similarly, the superfield $\bf Z$ of the phase-symmetric case (Subsection \ref{subsec_phase}) can be identified with the composite superfield $\bf P$ of Eq.~\eqref{P_def} to obtain analogous redefinitions in terms of the components of $\bf Z$.

The superfield $\bf T$ constrained by \eqref{constr_XDT} and \eqref{constr_XTT} also satisfies the secondary constraint $({\bf T}+\overbar{\bf T})^3=0$. But it should be noted that the composite scalar $t={\rm Re}T$ given by \eqref{t_sol} does not satisfy our solution \eqref{Sigma_sol} to such cubic constraint because it was obtained under the assumption of non-vanishing $F$-term, while the auxiliary field $F^t$ vanishes at the vacuum because it starts with the goldstino and spacetime derivatives (see Eq.~\eqref{Ft_sol}).

\section{Conclusion}\label{sec_concl}

In this work we found non-trivial solutions to real cubic constraints on chiral superfields transforming under global $U(1)$ symmetries. For a chiral superfield $\bf S$ transforming by an imaginary shift we considered the invariant constraint $({\bf S}+\overbar{\bf S}-2a)^3=0$, which is solved by Eq.~\eqref{phi_sol} -- this eliminates the real part of the scalar $S\equiv {\bf S}|$ in terms of the goldstino and the axion which is the imaginary part of $S$. For a chiral superfield $\bf Z$ transforming under the $U(1)$ by a phase we impose the constraint $(|{\bf Z}|^2-b^2)^3=0$, which is solved by Eq.~\eqref{z_sol} assuming $b\neq 0$. In this case the radial mode of $Z=|Z|e^{i\zeta}$ is eliminated, while the axion is identified with its angular part $\zeta$. We showed that the cubic constraints are generalizations of the corresponding quadratic constraints $({\bf S}-a)^2=0$ and $({\bf Z}-b)^2=0$. In particular, each of the cubic constraints admits two solutions -- one that eliminates saxion only (which is our main result), and one that eliminates both axion and saxion and satisfies the corresponding quadratic constraint as well. Along the way we proved that the constraints $(|{\bf Z}|^2-b^2)^3=0$ and $[\log(|{\bf Z}|^2/b^2)]^3=0$ lead to the same solution.

We constructed a minimal Lagrangian for the shift-symmetric cubic constraint, where the goldstino-axion interactions can be seen. The full solution to the $F$-term equation of motion as well as the full on-shell Lagrangian can be found in Appendix \ref{Appendix_F_term}.

Finally, we compared our results to the Komargodski--Seiberg \cite{Komargodski:2009rz} model of effective goldstino-axion dynamics, which uses the well-known quadratic constraint for the goldstino chiral superfield, ${\bf X}^2=0$, together with the axion chiral superfield $\bf T$ satisfying the orthogonality constraint ${\bf X}({\bf T}+\overbar{\bf T})=0$, so that only the imaginary scalar component of $\bf T$ survives. We showed that the independent fields in our approach (axion, goldstino, and $F$, all in the same chiral superfield) can be related to the independent fields of the KS model (goldstino and $F$-term in $\bf X$, and axion in $\bf T$) by non-linear field redefinitions. There are two notable differences between the two approaches. First, our approach uses a single independent chiral superfield satisfying a single constraint (not counting the chirality constraint), whereas in the KS approach we have two chiral superfields and two constraints. And second, solving the cubic constraints leads to the terms in the Lagrangian with negative powers of ($|F|^2-\partial\varphi\partial\varphi$), which explicitly shows the upper bound on the kinetic energy of the axion (in the KS model this upper bound is implicit) and suppresses higher-derivative terms in the Lagrangian, acting as a cut-off scale and effectively SUSY breaking scale. To be specific, the upper bound on the axion kinetic energy applies to its spatial derivatives and not temporal (we use ``mostly plus'' metric signature).

It should also be noted that although the two approaches can be related by non-linear field transformations, they do not necessarily represent the same physics since the actions have different structures. For example the aforementioned suppression scale ($|F|^2-\partial\varphi\partial\varphi$) is directly affected by the derivatives of $\varphi$, in contrast to the KS model.

In future works it would be natural to generalize our results to local abelian symmetries, as well as local supersymmetry. On the other hand our cubic constraints can be applied in axion phenomenology, both in cosmology and particle physics, similarly to orthogonal nilpotent superfields but with possibly different implications.~\footnote{After our paper was posted on arXiv, Ref. \cite{Terada:2021rtp} appeared where the author constructed minimal supergravity inflation using our cubic nilpotent superfields, and showed that the resulting theory does not suffer from catastrophic gravitino production due to vanishing sound speed, unlike theories using orthogonal nilpotent superfields.}

\section*{Acknowledgements}

This work is supported by CUniverse research promotion project of Chulalongkorn University (grant CUAASC). A.C. was supported by ``CU Global Partnership Project" under the grant No. B16F630071.

\begin{appendices}
\numberwithin{equation}{section}

\section{Off-shell SUSY transformations}\label{App_SUSY}

Components of a (free) chiral superfield,
\begin{equation}
    {\bf S}=S+\sqrt{2}\theta\chi+\theta^2F~,
\end{equation}
transform under $N=1$ SUSY as,
\begin{equation}
    \delta_\epsilon S=\epsilon\chi~,~~~\delta_\epsilon\chi_\alpha=i\sigma^m_{\alpha\dot\alpha}\bar\epsilon^{\dot\alpha}\partial_m S+\epsilon_\alpha F~,~~~\delta_\epsilon F=i\bar\epsilon\bar\sigma^m\partial_m\chi~.
\end{equation}

For the shift-symmetric case of Subsection \ref{subsec_shift}, taking $S=\phi+i\varphi$, and eliminating $\phi$ according to the cubic constraint \eqref{cubic_shift_constr} with the solution \eqref{phi_sol}, we can obtain the transformation rules for the independent fields $\varphi$, $\chi$, and $F$. The transformation of $\varphi$ and $F$ are unaffected by the cubic constraint, where for $\varphi$ we have,
\begin{equation}
    \delta_\epsilon\varphi=-\tfrac{i}{2}(\epsilon\chi-\bar\epsilon\bar\chi)~.
\end{equation}
The transformation of the goldstino $\chi$, however, contains $\partial_m\phi$ and after $\phi$ is eliminated, becomes a complicated function involving $U\equiv 2(|F|^2-\partial_m\varphi\partial^m\varphi)$ and $\beta$ given by Eq.~\eqref{beta_def}. Up to the terms bilinear in $\chi$ and $\bar\chi$ we can write,
\begin{align}
    \delta_\epsilon\chi_\alpha &=\epsilon_\alpha F+\frac{i}{U}\sigma^m_{\alpha\dot\alpha}\bar\epsilon^{\dot\alpha}\left(iU\partial_m\varphi+\overbar F\chi\partial_m\chi+F\bar\chi\partial_m\bar\chi+\partial_n\varphi\partial_m\chi\sigma^n\bar\chi+\partial_n\varphi\chi\sigma^n\partial_m\bar\chi\right) \nonumber\\
    &\hspace{22em}+{\cal O}(\chi^2,\bar\chi^2,\chi\sigma^m\bar\chi)~.
\end{align}

Next we consider the phase-symmetric case of Subsection \ref{subsec_phase}. Among the components of the chiral superfield ${\bf Z}$  we eliminate the radial scalar $|Z|$ according to the solution \eqref{z_sol}. This leaves the transformation of the auxiliary field untouched, $\delta_\epsilon F^z=i\bar\epsilon\bar\sigma^m\partial_m\psi$, while the axion and the goldstino transform as, 
\begin{align}
    \delta_\epsilon\zeta &=-\frac{i}{2}\delta_\epsilon\log\frac{Z}{\overbar{Z}}=-\frac{i}{2b}\left(e^{-i\zeta}\epsilon\psi-e^{i\zeta}\bar\epsilon\bar\psi\right)+{\cal O}(\psi^2,\bar\psi^2)~,\\
    \delta_\epsilon\psi_\alpha &=\epsilon_\alpha F^z+\frac{i}{Y}\sigma^m_{\alpha\dot\alpha}\bar\epsilon^{\dot\alpha}e^{i\zeta}\left(ibY\partial_m\zeta+\overbar F^ze^{-i\zeta}\psi\partial_m\psi+F^ze^{i\zeta}\bar\psi\partial_m\bar\psi\right.\nonumber\\
    &\qquad\qquad\qquad\left.+b\partial_n\zeta\partial_m\psi\sigma^n\bar\psi+b\partial_n\zeta\psi\sigma^n\partial_m\bar\psi\right)+{\cal O}(\psi^2,\bar\psi^2,\psi\sigma^m\bar\psi)~,
\end{align}
where $Y\equiv 2(|F^z|^2-b^2\partial_m\zeta\partial^m\zeta)$.

\section{Integrating out the \texorpdfstring{$F$}{}-term}\label{Appendix_F_term}
Here we integrate out the $F$-term in the model of Section \ref{sec_eg}. The Lagrangian reads,
\begin{align}
    {\cal L}&=\int\!d^4\theta\,{\bf S}{\bf\bar S}+\left(\mu\int\!d^2\theta\,{\bf S}+{\rm h.c.}\right)\nonumber\\
    &=\phi\Box\phi+\varphi\Box\varphi-i\chi\sigma^m\partial_m\bar\chi+\mu(F+\overbar F)+|F|^2~,\label{App_L}
\end{align}
where we used $S=\phi+i\varphi$, and chose the parameter $\mu$ to be real.

Applying the cubic shift-symmetric constraint \eqref{cubic_shift_constr}, we use its solution \eqref{phi_sol} to eliminate $\phi$ as a function of the goldstino $\chi,\bar\chi$, its derivatives, and the derivative of the axion $\partial_m\varphi$. In this case it is most convenient to work with the form of the solution given by Eq.~\eqref{Sigma_sol}, so that $\phi$ becomes,
\begin{equation}
    \phi=\frac{B}{2A}-\frac{B^2}{8A^4}\Box B~,
\end{equation}
where we set $a=0$ since it is irrelevant for the Lagrangian. Let us recall the definitions of $A$ and $B$,
\begin{eqnarray}
	A&\equiv & 2|F|^2-2\partial_m\varphi\partial^m\varphi-i\chi\sigma^m\partial_m\bar{\chi}+i\partial_m\chi\sigma^m\bar{\chi}~,\\
	B&\equiv &\overbar{F}\chi^2+F\bar{\chi}^2+2\partial_m\varphi\chi\sigma^m\bar{\chi}~.
\end{eqnarray}
Then the term $\phi\Box\phi$ can be written as,
\begin{align}
    \phi\Box\phi=\frac{B\Box B}{4A^2}-\frac{B}{2A^3}\partial_mA\partial^mB&-\frac{B\Box B}{16A^4}\Box(\chi^2\bar\chi^2)\nonumber\\
    &+\frac{\chi^2\bar\chi^2}{4A^4}\left[2A\partial_mA\partial^mA-A^2\Box A-\tfrac{1}{4}(\Box B)^2\right]~,\label{App_phibox}
\end{align}
where we used the fact that $B^2=A\chi^2\bar\chi^2$. The first term of \eqref{App_phibox} is at least bilinear in $\chi$ and $\bar\chi$, the second term is at least trilinear, while the rest are quadrilinear terms.

Using the Lagrangian \eqref{App_L} we write down the Euler--Lagrange equation of $F$,
\begin{align}
    \left[\frac{\partial}{\partial F}-\partial_m\frac{\partial}{\partial(\partial_m F)}+\right.&\left.\partial_m\partial_n\frac{\partial}{\partial(\partial_m\partial_n F)}\right]{\cal L}\nonumber\\
    &=\mu+\overbar F+\left[\frac{\partial}{\partial F}-\partial_m\frac{\partial}{\partial(\partial_m F)}+\Box\frac{\partial}{\partial(\Box F)}\right]\phi\Box\phi=0~,\label{App_EL}
\end{align}
where we notice that the second derivatives of $F$ enter $\phi\Box\phi$ only as $\Box F$.

The process of solving the above equation can be simplified if we use the following ansatz for $\overbar F$,
\begin{equation}
    \overbar F=-\mu+\Gamma_1\chi^2+\Gamma_2\bar\chi^2+\Omega_m\chi\sigma^m\bar\chi+E~,\label{App_Fbar}
\end{equation}
where $\Gamma_1$, $\Gamma_2$, and $\Omega_m$ are some complex functions of $\partial_m\chi$, $\partial_m\bar\chi$, and $\partial_m\varphi$, while $E$ collectively denotes all the terms trilinear and quadrilinear in $\chi$ and $\bar\chi$. If we plug this ansatz into Lagrangian \eqref{App_L}, we can see that all the $E$- and $\overbar E$-dependent terms vanish due to the fact that $E$ is at least trilinear, and so $E\chi^2=E\bar\chi^2=E\chi\sigma^m\bar\chi=0$. 

Therefore we need to find only $\Gamma_1$, $\Gamma_2$, and $\Omega_m$, which can be done by plugging ansatz \eqref{App_Fbar} into Eq.~\eqref{App_EL} and keeping only the bilinear terms,
\begin{equation}
    \Gamma_1\chi^2+\Gamma_2\bar\chi^2+\Omega_m\chi\sigma^m\bar\chi=-\left.\left\{\left[\frac{\partial}{\partial F}-\partial_m\frac{\partial}{\partial(\partial_m F)}+\Box\frac{\partial}{\partial(\Box F)}\right]\phi\Box\phi\right\}\right|_{\rm bi}~.
\end{equation}
This yields,
\begin{align}
    \Gamma_1&=-\mu^2~\frac{\mu\partial_m\chi\partial^m\chi+\mu\partial_m\bar\chi\partial^m\bar\chi-2\partial_m\varphi\partial_k\chi\sigma^m\partial^k\bar\chi}{4(\mu^2-\partial_n\varphi\partial^n\varphi)^3}~,\\
    \Gamma_2&=-\partial_l\varphi\partial^l\varphi~\frac{\mu\partial_m\chi\partial^m\chi+\mu\partial_m\bar\chi\partial^m\bar\chi-2\partial_m\varphi\partial_k\chi\sigma^m\partial^k\bar\chi}{4(\mu^2-\partial_n\varphi\partial^n\varphi)^3}~,\\
    \Omega_m&=\tfrac{1}{2}(\mu^2-\partial_p\varphi\partial^p\varphi)^{-3}\Big[\mu\partial_m\varphi(\mu\partial_n\chi\partial^n\chi+\mu\partial_n\bar\chi\partial^n\bar\chi-2\partial_n\varphi\partial_k\chi\sigma^n\partial^k\bar\chi)\nonumber\\
    &\hspace{5em} -(\mu^2+\partial_q\varphi\partial^q\varphi)\eta_{mn}\partial_k\varphi\partial_l\bar\chi\bar\sigma^{nk}\partial^l\bar\chi-2\mu^2\eta_{mn}\partial_k\varphi\partial_l\chi\sigma^{nk}\partial^l\chi\Big]~,
\end{align}
so that $\Gamma_1$ and $\Gamma_2$ are real, whereas $\Omega_m$ stays complex.

Finally, the full Lagrangian after $F$ and $\overbar F$ are eliminated reads,
\begin{align}
    {\cal L}&=\varphi\Box\varphi-i\chi\sigma^m\partial_m\bar\chi-\mu^2+\chi^2\bar\chi^2\left(\Gamma_1^2+\Gamma_2^2-\tfrac{1}{2}\Omega_m\overbar\Omega^m\right)+\frac{{\cal B}\Box{\cal B}}{4{\cal A}^2}-\frac{{\cal B}}{2{\cal A}^3}\partial_m{\cal A}\partial^m{\cal B}\nonumber\\
    &\qquad+\frac{\chi^2\bar\chi^2}{4{\cal A}^4}\left[2{\cal A}\partial_m{\cal A}\partial^m{\cal A}-{\cal A}^2\Box{\cal A}-\tfrac{1}{4}(\Box{\cal B})^2\right]-\frac{{\cal B}\Box{\cal B}}{16{\cal A}^4}\Box(\chi^2\bar\chi^2)~,\label{App_L_final}
\end{align}
where we have defined,
\begin{align}
    {\cal B}&\equiv B|_{F}=-\mu(\chi^2+\bar\chi^2)+2\chi\sigma^m\bar\chi\partial_m\varphi+2\chi^2\bar\chi^2\Gamma_2~,\\
	{\cal A}&\equiv A|_{F,{\rm bi}}=2(\mu^2-\partial_m\varphi\partial^m\varphi)-i\chi\sigma^m\partial_m\bar{\chi}+i\partial_m\chi\sigma^m\bar{\chi}\nonumber\\
	&\qquad\qquad\quad~-2\mu\left[(\chi^2+\bar\chi^2)(\Gamma_1+\Gamma_2)+\chi\sigma^m\bar\chi(\Omega_m+\overbar\Omega_m)\right]~.
\end{align}
The notation $|_F$ means using solution \eqref{App_Fbar} for $F$ and $\overbar F$, and $|_{\rm bi}$ means extracting terms that are at most bilinear in the goldstino (higher order terms in $A$ will vanish when plugged into the Lagrangian).

\end{appendices}

\providecommand{\href}[2]{#2}\begingroup\raggedright\endgroup

\end{document}